# 96 dB Linear High Dynamic Range CAOS Spectrometer Demonstration

Mohsin A. Mazhar and Nabeel A. Riza, *Fellow, IEEE*

*Abstract*— For the first time, a CAOS (i.e., Coded Access Optical Sensor) spectrometer is demonstrated. The design implemented uses a reflective diffraction grating and a time-frequency CAOS mode operations Digital Micromirror Device (DMD) in combination with a large area point photo-detector to enable highly programmable linear High Dynamic Range (HDR) spectrometry. Experiments are conducted with a 2850 K color temperature light bulb source and visible band color bandpass and high-pass filters as well as neutral density (ND) attenuation filters. A ~369 nm to ~715 nm input light source spectrum is measured with a designed ~1 nm spectral resolution. Using the optical filters and different CAOS modes, namely, Code Division Multiple Access (CDMA), Frequency Modulation (FM)-CDMA and FM-Time Division Multiple Access (TDMA) modes, measured are improving spectrometer linear dynamic ranges of 28 dB, 50 dB, and 96.2 dB, respectively. Applications for the linear HDR CAOS spectrometer includes materials inspection in biomedicine, foods, forensics, and pharmaceuticals.

*Index Terms*— Spectrometer, Digital Micromirror Device, Spectroscopy, Optical MEMS, High Dynamic Range.

## I. INTRODUCTION

OPTICAL spectrometry is a powerful measurement tool for numerous applications. Spectrometers can be designed for single spatial point/zone inspection or for pixel-by-pixel image format spectroscopic mappings [1]. There are a variety of methodologies to extract spectral information. Given the maturity of silicon-based image sensors such as CCD and CMOS image sensors, their use as sensors in spectrometer has been dominant for UV, visible and Near Infrared (NIR) spectrometry between 350 nm and 1100 nm [2-6]. For wavelengths exceeding 1100 nm, other non-silicon [7] optical image photo-detectors (PD) technologies have been deployed at the expense of higher cost, limited dynamic range (DR) and greater nonlinearities. One solution to reduce costs and improve performance is the classic dispersive grating spectrometer design that uses a point PD and deploys a scanning slit or grating to time sequentially scan through the spectrum under measurement. A potential weakness of such a design is that only a small fraction of the light in the full spectrum is selected for photo-detection at any instant causing limitations in measured spectral bin Signal-to-Noise Ratio (SNR) leading to non-robust readings.

Over 70 years ago, an elegant solution to solve this low SNR problem was proposed that involved using many simultaneous spatially coded slits using patterned disks allowing all spectral bins (i.e., much more total light power) to be detected at the same time giving higher SNR photo-detection and hence a higher reliability spectral measurement [8]. This technique was again independently proposed and developed over the following decades by other researchers and eventually was called Hadamard Transform (HT) spectrometry given that the Hadamard orthogonal matrix content was used for both spatially encoding the spectral channels and then decoding the spectral bins via inverse HT [9-11]. Interestingly spinning digital spatial coding optical disks were also proposed and demonstrated for RF spectrum analysis [12]. Another remarkable innovation happened on the optical spatial light modulation device front when in the late 1990's, the TI DMD became available for use by researchers [13]. The DMD is a highly reliable digital on/off micromirror tilt states near 2 million pixels (micromirrors) broadband (350-2700 nm) spatial light modulator with frame rates currently reaching 50 KHz. Researchers realized the value of the DMD and proposed and demonstrated systems involving optical spectral control for a host of applications [14-19]. In fact, the DMD was also deployed to realize a HT spectrometer [20] and even recently studies continue on the HT DMD-based spectrometer and its higher speed compressive designs [21-24].

Use of speckle via an inserted scattering medium has been recently suggested to improve the spectral range DR of a dispersion prism spectrometer that uses a CCD sensor camera as the image sensor PD, thus restricting wider bandwidth spectral usage [25]. It would also be highly desirable that a dispersive-optic based point PD spectrometer not only has a programmable adequate SNR, but also delivers spectral measurements that are linear and with a detected signal HDR (> 90 dB). In addition, faster speeds than the classic HT spectrometer would be desirable. Recently the CAOS platform has been proposed and demonstrated [26] that allows the design of a SNR controllable linear extreme DR optical imager using the principles of the HDR multiaccess RF wireless phone network. CAOS exploits time-frequency coding of selected optical pixels along with low noise programmable gain coherent signal processing involving RF correlation and spectral processing to extract HDR optical pixel data. In addition, it has been pointed out that the CAOS platform when using its CDMA-mode [27] also allows the design of a faster speed spectrometer when compared to the HT spectrometer [28]. Hence, this letter presents the first demonstration of a linear HDR CAOS spectrometer. Specifically, a broadband light source is engaged with a DMD-based CAOS point PD grating spectrometer design to demonstrate linear HDR spectroscopy using ND optical filter calibrated color filter-based spectroscopic test targets. Described are the details of the CAOS spectrometer design, experiment and sample test results.

## II. CAOS SPECTROMETER

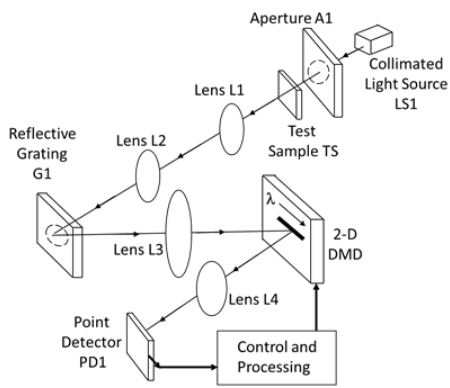

Fig. 1 3-D view of the linear HDR CAOS spectrometer.

Fig.1 shows the 3-D view of a basic CAOS spectrometer design. A collimated beam from an external light source LS1 passes through iris aperture A1 and test sample TS to enter a telescopic lens system comprised of spherical lenses L1 and L2 with focal lengths F1 and F2, respectively. The aperture size is controlled to illuminate the reflective grating G1 with the appropriate spot beam size given a larger beam size improves the system optical efficiency and diffraction limited spectral resolution while a smaller beam size improves the collimated beam illumination zone given the light source provides a partial spatially collimated beam. The surface normal of the grating makes an angle α with the beam optical axis and is adjusted for efficient placement of the desired spectrum on the DMD surface. After passing through the focus lens L3 with focal length F3, the spatially dispersed light from the grating representing the incident light spectrum forms a line illumination along the DMD horizontal axis labelled as λ. The +1 (i.e., +θ) tilt micro-mirror state of the DMD sends light from the DMD plane through the imaging lens L4 of focal length F4 to be imaged on to a large area high speed point detector PD1. The photo-detected signal is amplified and sampled by a high speed Analog-to-Digital Converter (ADC) and then sent for time-frequency (Hz domain) Digital Signal Processing (DSP) operations via the control and signal processing electronics. Depending on the required linear DR needs of the spectral testing scenario, the CAOS spectrometer hardware is programmed to operate with the optimal time-frequency CAOS mode [27].

The basic CAOS-mode is the CDMA mode where each CAOS pixel chosen in the DMD micromirrors zone is assigned a unique binary code sequence in time selected ideally from an orthogonal binary code set such as from a Walsh code set. The location of these M CAOS pixels can be anywhere in the DMD optical spectral 1-D λ grid space and the size and shape of the CAOS pixel in micromirror units is also programmable based on spectral resolution needs for the spectroscopic sample under test. In effect, the shortest time code sequences can be used in the CDMA-mode as only the few select CAOS pixels of spectral interest are harvested by the CAOS spectrometer making a higher speed spectrometer versus for example the HT spectrometer that implements continuous spatial coded space HTs [28]. Given all CAOS pixels of interest are photo-detected at the same time, higher light levels for photo-detection can lead to adequate SNR (i.e., >1) and adequate (e.g., 30 dB) dynamic range spectral bin measurements. For CDMA-mode operations, light source power levels as well as photo-detector optical gain and electronic amplifier gain and filter settings in electronic hardware can be adjusted for optimal SNR spectral bin measurements. To extract spectral data over a higher linear DR, the FM-CDMA CAOS-mode can be deployed that not only provides a higher DR but also software-based SNR control via DSP gain adjustments. Specially, each CDMA code bit at a bit rate of $f_B$ bits/sec in the CAOS pixel time code sequence is modulated by an FM carrier at $f_c$ Hz where $f_C = f_B$ P and P is an integer. Recovery of the irradiance for a CAOS optical spectral bin is achieved in two decoding steps where the first step involves RF spectral processing of the FM carrier within each bit time via the DSP Fast Fourier Transform (FFT) algorithm to recover the irradiance of all the selected CAOS spectral bins and the second step involves decoding of these various simultaneous spectral bins that were encoded with CDMA codes. By controlling the ADC $f_S$ sampling rate, $f_c$ and $f_B$, the FFT DSP gain of $10\log(N/2)$ dB is controlled by controlling the number of data samples N for the N-point FFT processing. For lower noise and higher DR spectral bin recovery, both the CAOS spectral bins M and the sample count N per bit time can be optimized based on prior knowledge of sample spectral test data. To enable extreme linear DR, the FM-TDMA CAOS mode can be engaged where a single CAOS pixel is extracted per TDMA time slot, allowing optimal use of the full saturation capacity of the point PD. Previous data measured for a CAOS camera indeed highlighted these controllable SNR and linear DR features of the various CAOS modes and in fact have shown a 177 dB linear extreme DR capability for the imaging application [29]. The next section highlights similar linear HDR capabilities for the CAOS spectrometer instrument.

### III. EXPERIMENTS AND DISCUSSION

The Fig.1 CAOS spectrometer is built in the laboratory using the following components. LS1: Avantes (UK) AvaLIGHT-HAL-S-Mini Pro-lite 2850 K bulb color temperature, 4.5 mW power, 360 – 2500 nm spectrum, 600 micron diameter fiber feed with a 5 cm focal length 2.5 cm diameter collimation lens with a 2.1 degree output beam divergence; Vialux (Germany) DMD model V-7001 with micromirror size of 13.68 µm x 13.68 µm; DELL 5480 Latitude laptop for control and DSP, National Instruments 16-bit ADC model 6211; Thorlabs components that include test sample 450 nm and 620 nm color 10 nm FWHM bandpass filters FB620-10 and FB450-10 and a red dichroic band high-pass filter FD1R and a set of ND filters, point PMT Model PMM02 (and optional silicon point PD model PDA100A2 with 70 dB variable electronic gain amplifier), 4 mm diameter Iris A1, 5.08 cm diameter uncoated broadband lenses L1/L2/L3 with F1=5 cm, F2=F3=6 cm, and a spherical mirror SM1 with F4=3.81 cm focal length used instead of the lens L4, G1 with $f_G$=600 lines/mm and $D_g$=1.62 nm/mrad dispersion at 750 nm. Key inter-component distances are: 11 cm between L1 and L2 giving an on-axis 6 mm diameter beam at the G1 plane with G1 rotated with incidence angle $\alpha$ = 6 degrees. 21.2 cm between A1 and G1, 10 cm between DMD and SM1/L4 and 6.2 cm between SM1/L4 and point PD1.

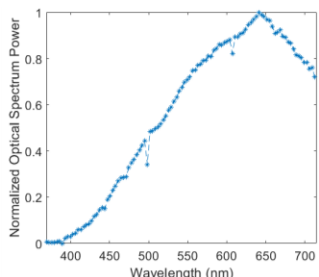

Fig. 2 Spectrometer full spectral system response via CAOS CDMA-mode.

Fig.2 shows the experimental CAOS spectrometer system optical spectrum measured by the CAOS CDMA-mode using W=128 bits length Walsh codes at a $f_B$=1 KHz bit rate and without placing a TS. A spectrum of ~369 nm to ~715 nm is observed over the full width of the DMD using M=102 CAOS pixels (3.4 nm/CAOS pixel) or ~0.34 nm/micromirror as each CAOS pixel has a width of 10 micromirrors and a height of 300 micromirrors. The spectrometer wavelength axis is coarsely calibrated by placing known wavelength narrow 10 nm FWHM bandpass color filters as TSs such as at 450 nm, 550 nm and 620 nm. Note that a CAOS pixel width of 3 micromirrors also allowed a full spectrum recovery indicating a system spectral resolution of 0.34 x 3=1.02 nm. Ideally, a fine electronically tuned calibrated optical filter should be used to calibrate the CAOS spectrometer wavelength axis on the DMD. Using high-pass filter TS, Fig.3 shows the measured spectrum after normalization using the inverse of the Fig.2 measured non-uniform system spectrum. At 550 nm, the deployed high-pass filter has a designed filter spectral gain of ~zero, as also closely confirmed by the Fig.3 spectral data at the 552.8 nm data point. This filter also has a rippled variable gain design near its rising band edge and this behavior is also confirmed in the Fig.3 data with a first maximum at 607.1 nm. Note that other maximum response points also occur, including at the 631 nm data point.

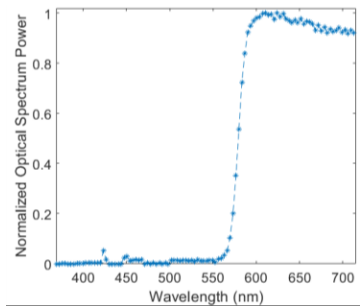

Fig. 3 Red color high-pass filter TS spectrum via CAOS CDMA-mode.

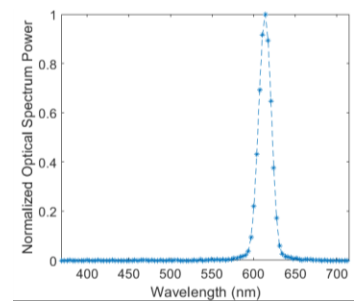

Fig.4 620 nm bandpass filter at OD=2.5 TS spectrum via FM-CDMA-mode.

The key feature provided by the CAOS spectrometer is programmable HDR and to test this aspect, the known 620 nm bandpass filter is used as a TS along with ND filters to attenuate the spectrum. Experiments show a robust test spectrum recovery for DR values of 28 dB and 50 dB for CDMA and FM-CDMA modes, respectively. Fig.4 shows the measured spectrum at the ND optical density (OD) filter value of 2.5 (i.e., 50 dB attenuation in the DR range) for the FM-CDMA mode using an FM of $f_c$=520 Hz, W=128 bits CDMA Walsh codes (via a 128 x128 Sylvester construction Hadamard matrix) at $f_B$=0.5 Hz bit rate (2 sec bit time) giving P= $f_c/f_B$=1040 and an ADC sampling rate of 65535 Sps giving N=131072 for N-point FFT gain of 48 dB. Fig.4 shows the measured spectrum using the 620 nm bandpass filter TS with a measured 16.3 nm FWHM versus the ideal 10 nm given the non-optimal optical design of the system that does not minimize and compensate image blur and optical aberrations such as by custom techniques [30]. In addition, the system is designed using paraxial (i.e., near on-axis) rays, but non-paraxial conditions apply as the LS1 beam is not ideally collimated and has a 2.1 degree divergence that gives a near 14 mm spread for the $\Delta\lambda$ = 346 nm full spectrum light on the DMD versus a ~12.5 mm width using the grating frequency-based spectral width paraxial limit design expression of $f_G \Delta\lambda F$ where F=F3 for focus lens L3.

To improve the spectrometer recovery DR beyond 50 dB DR, stronger attenuation ND filters are applied and the FM-TDMA

mode is engaged for a larger CAOS pixel covering the null-to-null 620 nm spectrum with FM $f_c$=520 Hz and ADC sampling TDMA one time slot duration of $T_D$=2 sec. Again a 48 dB FFT gain is used and robust spectral recovery is achieved up-to OD=4.8 giving a 96.2 dB measured HDR with an SNR=1.2. Fig.5 shows the high linearity plot of the 96.2 dB HDR response of the CAOS spectrometer during the 620 nm filter TS test. Note that experiments with the point silicon PD as PD1 also produced HDR spectral results from the CAOS spectrometer, e.g., HDR= 80 dB using the CAOS FM-TDMA mode.

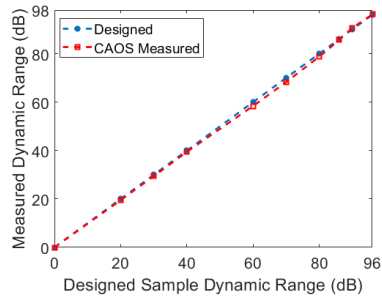

Fig. 5 Linearity plot for the 96 dB HDR response of the CAOS spectrometer.

To summarize the experimental results, first note that the reported wavelength range is only guaranteed when this spectrometer is compared with spectral measurements done with a commercial grade reference spectrometer. Also note that the spectrometry processing time is dominated by the deployed CAOS mode's encoding time $T_E$ that depends on the chosen CAOS-mode experimental parameters. For the Fig.2 & 3 CDMA-mode data, $T_E$=W/$f_B$=128/(1 KHz)=0.128 sec while the Fig.4 FM-CDMA mode data uses a $T_E$=W/$f_B$=128/(0.5 Hz)= 256 sec. Using a fastest $f_c$=25 KHz, $T_E$ can be reduced to 5.33 sec. The Fig.5 FM-TDMA mode data uses a $T_E$=$T_D$= 2 sec for each CAOS spectral reading. The highest linear HDR measured for the spectrometer is 96 dB with a minimum optical signal detected power of 10.1 pW maintaining a SNR ≥1.2.

## IV. CONCLUSION

For the first time, the CAOS spectrometer is demonstrated. Experiments using known spectral response color and ND filter test samples highlight the robust highly programmable spectral recovery. Specifically, spectrum measurements by the CAOS spectrometer with increasing test sample DR values from 28 dB, 50 dB, and 96.2 dB are achieved using CAOS CDMA, FM-CDMA, and FM-TDMA modes, respectively. The proposed CAOS spectrometer can be optimized for a chosen application and spectral range, include IR bands exceeding 1100 nm where non-silicon (e.g., Germanium) point PDs can be deployed.